\documentclass[fleqn,usenatbib]{mnras}

\usepackage{newtxtext,newtxmath}

\usepackage[T1]{fontenc}

\DeclareRobustCommand{\VAN}[3]{#2}
\let\VANthebibliography\thebibliography
\def\thebibliography{\DeclareRobustCommand{\VAN}[3]{##3}\VANthebibliography}

\usepackage{graphicx}	
\usepackage{amsmath}	
\usepackage{xcolor}

\newcommand{\asec}{$^{\prime\prime}$}
\newcommand{\cm}{cm$^{-3}$}
\newcommand{\lmcorc}{J0624–6948\ }
\newcommand{\Msun}{M$_{\odot}$}

\title[ORCs as SNRs]{On Odd Radio Circles as Supernova Remnants}

\author[S. K. Sarbadhicary  et al.]{
Sumit K. Sarbadhicary,$^{1,2,3}$\thanks{E-mail: sarbadhicary.1@osu.edu}
Todd A.\ Thompson$^{1,2,3}$
Laura A.\ Lopez$^{2,3}$ and
Smita\ Mathur$^{2,3}$
\\
$^{1}$Department of Physics, The Ohio State University, Columbus, Ohio 43210, USA\\
$^{2}$Center for Cosmology \& Astro-Particle Physics, The Ohio State University, Columbus, Ohio 43210, USA\\
$^{3}$Department of Astronomy, The Ohio State University, Columbus, Ohio 43210, USA\\
}

\date{Accepted XXX. Received YYY; in original form ZZZ}

\pubyear{2015}

\begin{document}
\label{firstpage}
\pagerange{\pageref{firstpage}--\pageref{lastpage}}
\maketitle

\begin{abstract}
The origin of arcmin-sized Odd Radio Circles (ORCs) found in modern all-sky radio surveys remain uncertain, with explanations ranging from starburst/AGN-driven shocks to supernova remnants (SNRs) in the low-density ambient medium. Using well-calibrated radio light curve models, we assess the possibility that ORCs are radio SNRs evolving in low ambient densities. Our models imply that ORCs 1-5 and \lmcorc (near the LMC) as SNRs must be within 200 kpc and 100 kpc from the Sun respectively, given their observed flux densities and angular sizes. To be evolving in the circumgalactic medium of the Milky Way, our models require ORCs 1-5 to be ejecta-dominated SNRs within 50 kpc, evolving in ambient densities of $(0.2-1.2) \times 10^{-3}$ \cm. However, this is statistically unlikely because ORCs 1-5 would have ages $<640$ yrs, much smaller than their expected lifetimes of $\gtrsim$10$^5$ yrs at these densities, and because the low SN rate and steep profile of the stellar halo imply a negligible number of ORC-like SNRs within 50 kpc. The circumgalactic medium SNR scenario for \lmcorc is more likely (though still low probability) compared to ORCs 1-5, as our models allow \lmcorc to be $\lesssim$3000 yrs. On the other hand, the interpretation of \lmcorc as a Sedov-Taylor SNR at 50 kpc (LMC) distance is possible for a wide range of ambient densities ($6 \times 10^{-4} - 0.5$ \cm) and ages $\sim$$(0.2-2.6) \times 10^4$ yr, while also being consistent with the local HI environment.

\end{abstract}




\section{Introduction}

Recently, a class of low radio surface-brightness, circular-shaped objects has been discovered by the Australian Square Kilometer Pathfinder Array (ASKAP) Evolutionary Map of the Universe (EMU) survey \citep{NorrisOriginal, NorrisReview, LMCORC}, the Giant Meterwave Radio Telescope \citep[GMRT,][]{Koribalski2021} and the Low Frequency Array \citep[LOFAR,][]{ORCLOFAR}. These objects, named ORCs 1-5\footnote{We restrict our sample to the ORCs detected at $\sim 1$ GHz, so we leave out the LOFAR ORC \citep{ORCLOFAR} from our analysis for consistency. However, as \cite{ORCLOFAR} mentioned, the object is an arcmin across and would have a 1 GHz flux density of 4.6 mJy assuming a spectral index $=-1$, similar to ORCs 1-5.} are between 1-3 arcmins in size, spanning integrated radio flux densities of 0.1-7 mJy at 1 GHz. Observations of ORCs thus far have not revealed any corresponding detection in X-ray or optical emission lines, preventing distance measurements and potentially confirming the origin of the structures. Three of the ORCs are coincident with an galaxy at redshifts of 0.2-0.6 in the center, raising the possibility that they are produced by few-hundred kpc-scale shocks driven by a starburst/AGN \citep{NorrisMeerkat} or galactic mergers \citep{Dolag2022}.

Since the morphology of ORCs closely resemble  supernova remnants (SNRs), but without any X-ray or optical emission, we are motivated to consider whether a subset of the observed objects may be SNRs in the extended halo of the Galaxy. The majority of SNRs in the Galaxy and the Magellanic Clouds are observable or were discovered in the radio band \citep{Green2019,Badenes2010, Bozzetto2017}. \cite{NorrisOriginal} found that the latitude distribution of the ORCs are higher than that of known SNRs in the Galaxy from the \cite{Green2019} catalog. The Galactic SNR population, however, is a heterogenous compilation of SNRs from various different surveys, the majority of which were restricted to the Galactic plane.

\cite{Omar2022} recently argued that ORCs could be SNRs  evolving in the low-density intragroup medium. \cite{LMCORC} found an object (called J0624–6948) that bears similarity to ORCs and is located in the outskirts of the LMC, about 3$^{\circ}$ from the eastern edge of LMC's radio continuum region (3$^{\circ}\sim$ 2.6~kpc if the object is at the 50~kpc distance of the LMC: \citealt{Pie19}). Given its lack of any central galaxy, steep spectral index, and large angular size, \cite{LMCORC} preferred to interpret \lmcorc as an intergalactic Type Ia SNR produced by a progenitor in the old LMC stellar halo, possibly the first discovered object of its kind. Therefore, like other phenomena in astrophysics, it is possible that ORCs are composed of two diverse populations, with one being relatively nearby SNRs in the extended Milky Way halo and another being at extragalactic distances.

Meanwhile, more sensitive, contemporary radio surveys have been increasingly discovering new SNRs \citep[e.g.][]{Anderson2017, Hurley-Walker2019, Heywood2022}, highlighting the incompleteness in the Galactic SNR population due to selection effects from different sensitivity limits, source confusion, and surveyed region of the sky \citep{Green2019}. Analysis of the luminosity function of SNRs in external galaxies, where distances, interstellar medium (ISM), and stellar population environments around SNRs are well characterized, have shown that a non-negligible fraction of SNRs are missing, and many of them could be of Type Ia origin evolving in low-density ISM \citep{Chomiuk09, S17}. Several new SNRs have been discovered in recent surveys covering higher galactic latitudes, with properties consistent with evolution in the low-density halo of the Milky Way \citep{Fesen2020, Fesen2021, Becker2021, Churasov2021, Churasov2022, Araya2022}. 
SN Ia progenitors are expected to be abundant in the old stellar disk and halo that extend beyond 1 kpc height above the midplane \citep{Mateu2018}. In distant galaxies, Type Ia SNe have been observed to explode at scale heights of $\gtrsim 1$ kpc \citep{Hakobyan2017}, and a small fraction even appear to be going off in the intergalactic/cluster medium \citep{GalYam2003, Maoz2005, Sand2011, Graham2015}. If ORCs are indeed a new variety of SNRs, they can serve as novel probes of the circumgalactic medium (CGM) environment, shock physics in hot tenuous media, and SN progenitors in remote, low-metallicity stellar populations. 

In this paper, we use the radio light curve models of \cite{S17} (hereafter, S17) to assess the possibility that the measured flux densities and sizes of ORCs 1-5 and \lmcorc are consistent with SNRs evolving in a low-density, halo medium. The high Galactic latitudes of the ORCs (favoring heights well above the Galactic midplane), the projected location of \lmcorc away from the main LMC disk, and the lack of X-ray and optical emission from any of these objects suggest a hot, low-density environment. These make the ORCs distinct from the Galactic SNRs near the plane, and therefore models of synchrotron emission from SNRs as a function of ambient medium density are needed to properly characterize these objects. The S17 models have been used previously to reproduce observations of extra-galactic and Galactic SNRs in a diversity of environments \citep[S17,][]{ Russell2020} as well as young Type Ia SNRs evolving in the warm/hot interstellar medium, such as G1.9+0.3 and SN 1885A \citep{S19}.

The paper is organized as follows. In Section 2, we summarize the the physics of the S17 light curve model, and assumptions about model parameters. Section 3 describes the evolution of the light curves at different values of the ambient density and electron acceleration efficiency. We then assess whether these light curves can reasonably reproduce the observed properties of ORCs 1-5 and \lmcorc.

\section{Method}
Radio detection of SNRs depends sensitively on the ambient density, explosion energy, age, properties of the accelerated electrons producing synchrotron radiation, and the SN rate, which in turn depends on the stellar mass, age, and metallicity distributions. In S17, we developed a model of SNR radio light curves to interpret SNR populations evolving in different environments with different explosion properties and used it to understand the properties of SNRs in M33. Here, we use this model to understand under what conditions and for what parameters radio SNRs may explain ORCs.

\subsection{Radio Light Curve Model of SNRs} \label{subsec:model}
The S17 model predicts the synchrotron radio luminosity of SNRs by combining the synchrotron radio emission models of \cite{Chevalier98} with the self-similar non-radiative shock models of \cite{Truelove99}. The radio emission model assumes that radio emission is synchrotron radiation produced by electrons accelerated in the turbulent magnetic field amplified by the SN shock \citep{Fermi1949, Bell1978, Blandford1978}. The S17 radio luminosity models reproduce the radio luminosity function and sizes of SNRs in M33 \citep{Gordon99, Chomiuk09} and are consistent with the  luminosity-sizes of Galactic SNRs from \cite{CaseBhat98}. The model was also applied to individual SNRs, including SN1885A and G1.9+0.3 \citep{S19} and other extragalactic populations of SNRs \citep[e.g. in M83,][]{Russell2020}. The analytical nature of the S17 model is ideal for scanning the wide parameter space of SNR shocks with observations, and it was verified by \cite{Truelove99} and later studies, such as \cite{Tang2017}, that these smooth solutions connecting the ejecta-dominated phase through Sedov-Taylor evolution are consistent with numerical results to within a few percent. 

Qualitatively, the S17 model predicts that the radio luminosity of a SNR initially increases in the ejecta-dominated stage and then decreases in the Sedov-Taylor stage (Figure 2 in S17). The initial increase is due to the emitting volume increasing faster than the rate of deceleration (which sets the magnetic field and cosmic ray electron energy densities). In the Sedov-Taylor stage, the shock experiences stronger deceleration due to the swept-up ISM, causing the deceleration effects to dominate and the radio luminosity to decrease. Higher radio luminosities are expected for greater ISM densities and ejecta energies, and larger remnants with longer lifetimes are expected for lower ISM densities and higher ejecta energies. These results are consistent with numerical investigations of the radio emission from SNRs \citep[e.g.][]{BV04, Pavlovic2018}.

\subsection{Model assumptions} \label{sec:modelassumptions}

We assume a standard 1.4 M$_{\odot}$ ejecta mass, a 10$^{51}$ erg Type Ia explosion with an ejecta density profile of $\rho_{ej}(v) \sim v^{-n}$ \citep[here $\rho_{ej}$ is the ejecta density, $v$ is the ejecta velocity, and the power-law index $n=10$ as assumed in S17 and][]{Chomiuk2012}, interacting with a uniform density ambient medium.  Type Ia SNe are associated with old stellar populations \citep{Maoz2014}, and their remnants are generally more symmetric than core-collapse SNRs in multi-wavelength images \citep{Lopez2011, Peters2013}. The remote locations and symmetric morphology of the ORCs can therefore be more easily explained with a Type Ia origin. In the case of \lmcorc, \cite{LMCORC} argued that a core-collapse origin is only possible with a high-velocity progenitor, with required velocities of 80-400 km/s at the distance of LMC to explain its current location. This scenario is statistically rare and also in tension with the symmetric morphology of the ORCs.

The light curve models depends on the assumed ambient density and temperature, magnetic field strength, parameters describing the amplified magnetic field and accelerated electron spectrum, observing frequency and the radio detection limit. We explore a broad range of parameters to see whether or not SNRs of the angular size, brightness, and frequency of ORCs can be produced in the halo of the Galaxy.

\subsubsection{Ambient Density and Temperature} \label{sec:ambdens}
We assume densities $n_0 = 10^{-1}-10^{-4}\ \mathrm{cm}^{-3}$ that are typical of the warm-hot phase of the interstellar/circumgalactic medium (CGM) \citep{Gupta2012} where we believe such SNRs may be evolving. For example, the ORC \lmcorc had an ISM density constrained to $\sim 7 \times 10^{-3}$ cm$^{-3}$ based on the size of the SNR, assuming a distance of 50 kpc and that the SNR is in the Sedov-Taylor stage \citep{LMCORC}. Such low densities are also consistent with the lack of any X-ray or forbidden line emission tracing the ORCs. Consistent with the symmetric morphology of the ORCs \citep{NorrisOriginal, NorrisMeerkat, LMCORC} as well as Type Ia SNRs \citep{Badenes2007, Lopez2011}, we assume that the ambient medium is homogeneous. 

We set the temperature of the surrounding medium to be $10^6$ K, similar to measurements of the Galactic CGM and the virial temperature of a Milky-Way-sized halo \citep{Mo2010, Das2019, Gupta2021, Martyenko2022}. While the supersonic SNR shock evolution does not directly depend on the temperature, the duration of the light curve will be affected by the larger sound speed and lower Mach-number shocks \citep{Tang05}. We explore this caveat further in Appendix \ref{sec:hotISM}, where we find that while the \cite{Tang05} model predicts a deviation away from the conventional Sedov-to-radiative transition in SNR shocks, the difference in the radio light curve evolution and lifetime of the SNR is not significant. We therefore continue using our default S17 model dynamics.

\subsubsection{Electron acceleration physics} \label{sec:electronphysics}
The radio luminosity depends on the physics of electron acceleration, which remains an active topic of research \citep{Caprioli2015, Marcowith2016}. \cite{Chevalier98} accounted for the energy in the magnetic field and accelerated electrons by parameterizing the fraction of the shock energy going into the amplified magnetic field as $\epsilon_b$, the fraction of shock energy going into the accelerated electrons as $\epsilon_e$ and the spectral index of the accelerated electron spectrum as $p$, i.e. $N(E) \propto E^{-p}$ for $E>m_e c^2$. 

Similar to most SN and SNR studies using the Chevalier et al. models, we assume these parameters to be constant, with the exception of $\epsilon_b$ where we take $\epsilon_b \propto \epsilon_{p}(v_s/c + 1/M_A)$ (S17), where $\epsilon_{p} \approx 0.1\%$ is the cosmic-ray proton acceleration efficiency \citep{Lacki2010}, $v_s$ is the shock velocity, and $M_A$ is the Alfv\'en Mach number. The relation follows results from particle-in-cell simulations from \cite{Caprioli2014a, Caprioli2014b} that show that magnetic fields are amplified in the shock precursor by streaming cosmic-ray protons initially driving non-resonant instabilities and later, resonant instabilities as the shock decelerates \citep{Bell1978, Bell2004}. 

The cosmic-ray spectrum has typically been observed to have $p\approx 2.2-2.4$ in young SNRs \citep[][and references therein]{Caprioli2012}, so we assume a median value of $p=2.3$. More recent simulations by \cite{Diesing2021} suggest that the spectrum will likely steepen at higher shock velocities, which may affect the shorter-duration ejecta-dominated phase in our light curves. We plan to incorporate this in future iterations of the S17 model. 

The $\epsilon_e$ parameter is typically the least constrained and understood. Values vary between $\epsilon_e = 0.1$ for SNe/GRBs \citep[][and references therein]{Chevalier2017} to $10^{-2} - 10^{-4}$ in young SNRs \citep[e.g.][]{Morlino2012, Reynolds2021}. In S17, we found that $\epsilon_e \sim 10^{-3} - 10^{-2}$ for the sample of SNRs in M33 that are mostly in the Sedov-Taylor stage. This is also consistent with expectations from the Galactic cosmic-ray spectrum where the fraction of shock energy in cosmic-ray protons $\epsilon_{p} \sim 0.1$ \citep[e.g.][]{Lacki2010} and the normalization ratio of the cosmic ray protons and electrons $K_{ep} \sim 10^{-2}$ \citep{Park2015}, so $\epsilon_e \sim K_{ep}\epsilon_p \sim 10^{-3}$. We assume for our purposes that $\epsilon_e = 10^{-3}-10^{-2}$, and we report results for variation in this range.

\subsubsection{Ambient Magnetic field} \label{sec:ambfield}
In S17, the luminosity depends on the amplified magnetic field $B \sim (8 \pi \epsilon_b \rho v_s^2)^{1/2}$, which is much larger than the ambient magnetic field ($B_0$). This relation however means that $B \rightarrow 0$ as $v_s \rightarrow 0$, so in S17 we set $B = \mathrm{max}(B, 4B_0)$ to ensure the post-shock field in the non-radiative stage is at least a simple compression of the ambient field, similar to the models of \cite{BV04} and observations of old SNRs \citep[e.g.][]{Loru2021}. This causes the synchrotron light curve to plateau in the late Sedov-Taylor stage because the effects of expansion and deceleration of the shock roughly cancel out \citep[Fig 2 in S17; also seen in the numerical simulations of][]{BV04}. This plateau luminosity in the late Sedov-Taylor stage becomes dependent on the assumed value of $B_0$.

The value of the ambient magnetic field $B_0$ in the halo of the Milky Way is uncertain. S17 assumed $B_0 \sim (9 \mu\mathrm{G}) n_0^{0.47}$, based on \cite{Crutcher1999} who showed from Zeeman splitting measurements of molecular clouds that $B_0 \sim n_0^{0.47}$, and global measurements of $B_0 \sim 5-9$ $\mu G$ from radio intensity and polarization maps of spiral galaxies \citep{Beck15}. While this dependence accounts for local variations in $B_0$ in the star-forming ISM, it implies that $B_0 \ll 1 \mu\mathrm{G}$ in the ambient Galaxy halo where $n_0$ may be $\lesssim 10^{-3}$ cm$^{-3}$. While star-forming galaxies  tend to have $1-10$\,kpc-scale radio haloes where the equipartition fields are of order $\mu G$ \citep[e.g.][]{Planck2016, CHANGES}, the value of $B_0$, on 100\,kpc scales is not directly known. As the supernova remnant slows its expansion,  the shock-amplified field weakens, and eventually the magnetic field energy density ($U_B$) will eventually becomes less than the energy density of the background cosmic microwave background radiation field ($U_{cmb}$),where cooling via inverse-Compton scattering will dominate over synchrotron cooling. Thus, the value of $B_0$ on 100\,kpc scales matters to the late-time radio luminosity of remnants. For the purposes of this paper, we set $U_B = U_{cmb}$, where $U_B = B^2/8\pi$ and $U_{cmb} = (4\sigma/c) T_{cmb}^{4}$ where $\sigma$ is the Stefan-Boltzmann constant, we get a lower limit on $B_0 = B_{low} \approx 2 \mu\mathrm{G} (T_{cmb}/2.7 \mathrm{K})^2$. We assume $B_{low} = 2\mu$G for $T_{cmb}=2.7$K, and therefore $B_0 = \mathrm{max}(B_0, B_{low})$.

\subsubsection{Detection Limit}
We assume that the SNR will remain detectable as long as it is 3$\sigma$ above the RMS noise of the survey. In this paper, we use the detection limits of the ASKAP survey of \cite{NorrisReview}, which reached an RMS = 30$\mu$Jy/bm, so the detection limit is about 90$\mu$Jy for objects smaller than the synthesized beam (or point-spread function) of 12\asec. The limit for objects larger than the beam size will scale as the number of beams spanned by the object, i.e. $\sqrt{D^2/\theta^2}$, where $D$ is the angular size of the object, and $\theta$ is the angular size of the beam. 

\begin{figure*} 
\centering
 \includegraphics[width=0.9\textwidth]{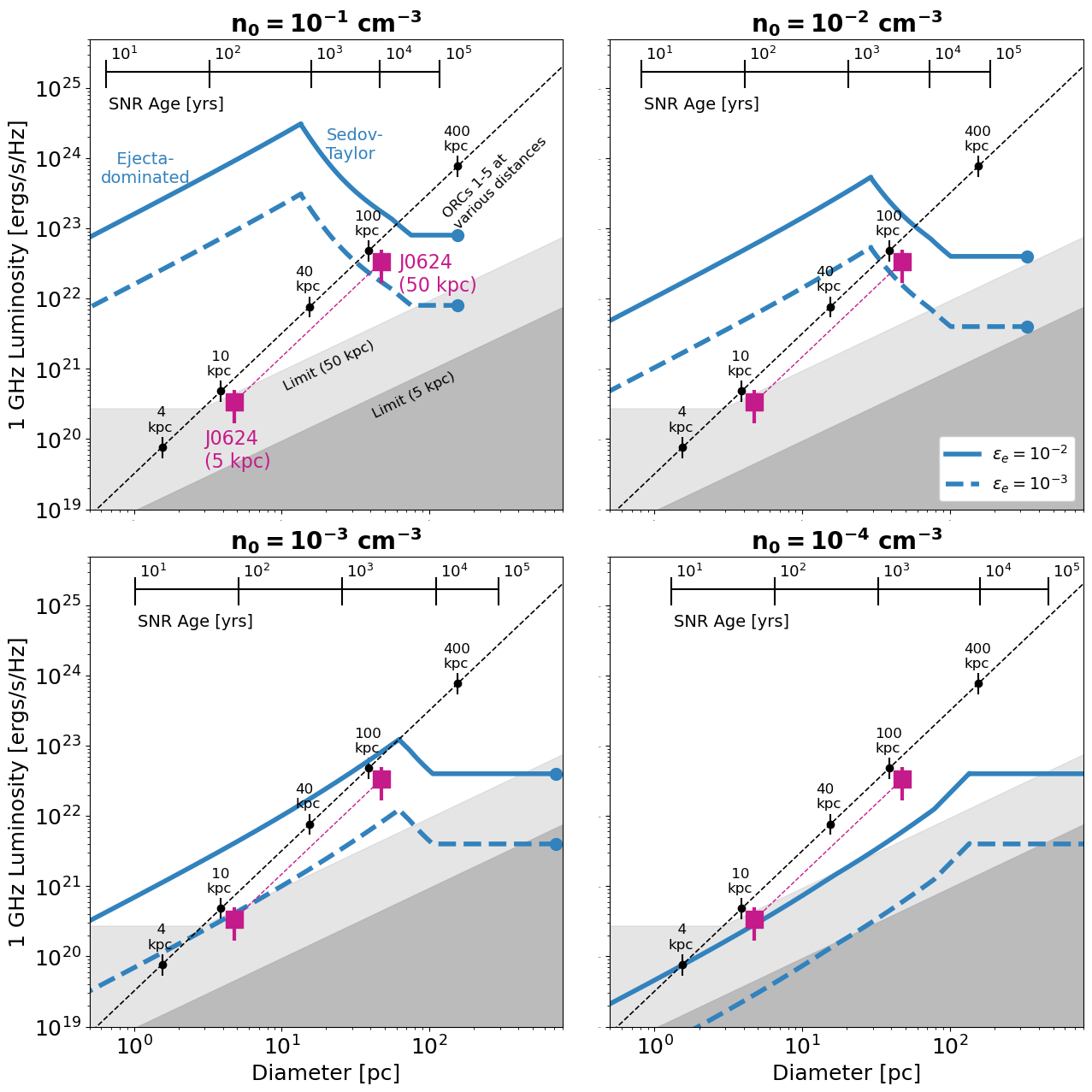}
 \caption{The 1 GHz SNR radio luminosity vs diameter predicted by the S17 model for $n_0$ = [$10^{-1}, 10^{-2}$, $10^{-3}$ , $10^{-4}$] cm$^{-3}$ (from top-left to bottom right) for different values of electron acceleration efficiency $\epsilon_e$ = [$10^{-2}$, $10^{-3}$] (solid, dashed blue lines). Horizontal scale with ticks represent predicted SNR ages for a given diameter. Blue circles at the end of the light curves mark the time when $M_s = 1$, i.e. the shock merges with the ISM. Shaded regions denote 3$\sigma$-detection limit (90$\mu$Jy/bm) of the EMU survey at 5 kpc (representative of the Milky Way Halo) and 50 kpc (representative of LMC). Purple square with error-bars represent \lmcorc at a distance of 5 kpc and 50 kpc. The diagonal dashed line with the vertical tick marks represent distances at which we assume ORCs 1-5 are located for the discussion in Section \ref{sec:orcs15}. The light curves indicate dimmer, smaller and longer lived SNRs with decreasing densities, and they would remain detectable above $3\sigma$ for the entirety of their lifetimes only for $n_0>10^{-3}$ \cm\ and $\epsilon_e > 10^{-3}$ at 5 kpc and $\epsilon_e > 10^{-2}$ at 50 kpc.} 
 \label{fig:lum3panel}
\end{figure*}

We adopt two characteristic distances for the SNRs, as in \cite{LMCORC}: 5\,kpc, representing the Milky Way stellar halo, and 50\,kpc, representing the LMC distance for assessing \lmcorc. 
At 5 kpc, the detection limit corresponds to a 1-GHz luminosity limit of $2 \times 10^{18}$ ergs/s/Hz and a beam physical size of 0.29 pc, while at 50 kpc (LMC), the luminosity limit is $2.7 \times 10^{20}$ ergs/s/Hz and the beam physical size is 2.9 pc. At these distances, most SNRs will be resolved in radio surveys, so their visibility will be surface-brightness-limited, as seen in Figure \ref{fig:lum3panel}.

\subsubsection{Observing Frequency} \label{sec:obsfreq}
We predict light curves at 1 GHz, roughly the frequency at which the ORCs were discovered \citep[e.g.]{NorrisOriginal, NorrisReview, LMCORC}. Both older SNRs \citep{Dubner2015} and the ORCs \citep{NorrisOriginal, LMCORC} are bright at these low frequencies because they have steep spectra ($\alpha \sim 0.5-0.6$) that are characteristic of non-thermal emission. Most modern synthesis arrays such as VLA, MeerKAT and ASKAP can observe in the 0.8-2 GHz frequency range. Historically, such low frequencies have been challenging to observe because of radio-frequency interference and dynamic range issues, but these have been substantially mitigated in modern surveys with sophisticated algorithms \citep[e.g.][]{Bhatnagar2011, Tasse2018}, making it feasible to obtain sensitive images where objects can be securely identified

\subsubsection{When does radio emission terminate?} \label{sec:whentoend}
When exactly radio emission stops in the SNR lifetime is difficult to model because of the poorly understood physics of electron acceleration. The SNR phase ends when the shock velocity becomes of the order of the sound speed of the ISM, but even before this stage, a cooling front will develop at the shock due to radiative losses \citep{Draine2011}. Additionally, cosmic ray acceleration becomes inefficient, and more cosmic rays escape from the post-shock region, as the shock slows \citep{Blasi2007, Xu2019}. As a result, models typically assume that radio emission stops at the onset of radiative phase \citep[e.g. S17, ][]{BV04}. 
However, at such high ambient temperatures (and possibly low metallicites if exploding in the halo), there is not a well-defined radiative phase as the internal energy is not radiated away before the SNR merges with the ISM \citep{Tang05, Karpov2020, Li2020}. We also note that radio emission from low-Mach number shocks ($M_s \sim 2$) has been observed in galaxy clusters and attributed to quasi-perpendicular shocks that preferentially accelerate electrons \citep{Guo2014, Xu2019}. 

We therefore conservatively assume that radio emission continues until the shock merges with the ambient medium (i.e. $M_s = 1$). 
Here $M_s = v_s/c_s$, where $v_s$ is the shock velocity, and $c_s = (0.12\ \mathrm{km/s}) \mu^{-1/2}T^{1/2}$ is the isothermal sound speed of the ambient medium. We assume $\mu=0.6$ and $T=10^{6}$ K for a hot, fully-ionized ambient medium, making $c_s = 155$ km/s.

\section{Results and Discussion}

\subsection{Radio properties of SNRs at low ambient densities}
We show our predicted 1-GHz luminosities and diameters as a function of SNR age for different ISM densities from $10^{-1}$ (top-left panel) to $10^{-4}$ \cm (bottom-right panel) for two different values of $\epsilon_e = 10^{-2}, 10^{-3}$ (solid, dashed) in Figure \ref{fig:lum3panel}. As mentioned in Section \ref{subsec:model}, the luminosity increases with age during the ejecta-dominated phase and decreases during the Sedov-Taylor phase. The radius monotonically increases with time, initially as $R_s \propto t^{0.7}$ for our assumed $\rho_{ej} \propto v^{-10}$ profile, and then $R_s \propto t^{0.4}$ during the Sedov-Taylor phase. The plateau portion in the late Sedov-Taylor phase is due to $B \rightarrow 4 B_0$, as explained in Section \ref{sec:ambfield}. Note that the plateau luminosity is the same for all panels since we assume the same $B_{low}$ (defined in Section \ref{sec:ambfield}). Overall, the SNR evolves more slowly as the ambient density decreases, as shown by the vertical age ticks shifting rightwards in Figure \ref{fig:lum3panel}. So for a given size, a SNR evolving in a lower density ambient medium will be younger than a higher density one. The radio emission is assumed to stop once the shock merges with the ISM, effectively ending the SNR phase. 

As seen in Figure \ref{fig:lum3panel}, for the current detection limits of ASKAP (marked by the gray shaded regions for 5 and 50 kpc), any SNR out to 5 kpc from the Sun will be detectable for the entirety of its lifetime, provided it is evolving in ambient densities $n_0 > 10^{-3}$ cm$^{-3}$, with $\epsilon_e \geq 10^{-3}$. Sources are brighter at a given radius when expanding into a higher density medium (except during the plateau phase set by B$_0$ in Section \ref{sec:ambfield}). Similarly, the more efficient the electron acceleration, the brighter the SNR. The size of the SNR is larger at lower densities for a given age ($R_s \propto n_0^{-1/5}$ during the Sedov-Taylor phase). The final size of the modeled SNRs, set by when $M_s \rightarrow 1$, is larger at low densities because $M_s \propto v_s \propto n_0^{-1/5}$ during the Sedov-Taylor phase). 
With decreasing density, the SNR becomes fainter and can drop below the detection limit of the survey. For example, at $n_0 = 10^{-4}$ cm$^{-3}$, the SNR will not be detectable at the distance of LMC, unless it has an $\epsilon_e > 10^{-3}$. For $n_0=10^{-2}$ cm$^{-3}$, SNRs are detectable throughout their lifetime at the distance of 5 kpc. At 50 kpc, SNRs are still detectable, though for $\epsilon_e < 10^{-2}$ they start to become surface brightness limited during the late Sedov phase.

To summarize, SNRs in the Milky Way halo, assuming a fiducial distance of 5 kpc, are likely to be detectable at $>3\sigma$ throughout their lifetime in surveys that can achieve EMU-like sensitivity, if they are evolving in ambient densities $\gtrsim 10^{-3}$ cm$^{-3}$ and $\epsilon_e > 10^{-3}$. The same is true for SNRs at the distance of LMC, but for $\epsilon_e \gtrsim 10^{-2}$. The detectability and visibility time noticeably decreases for lower densities, for larger distances, and for lower $\epsilon_e$ values. 

\subsection{Can ORCs 1-5 be SNRs?} \label{sec:orcs15}
\begin{figure}
    \includegraphics[width=1.13\columnwidth]{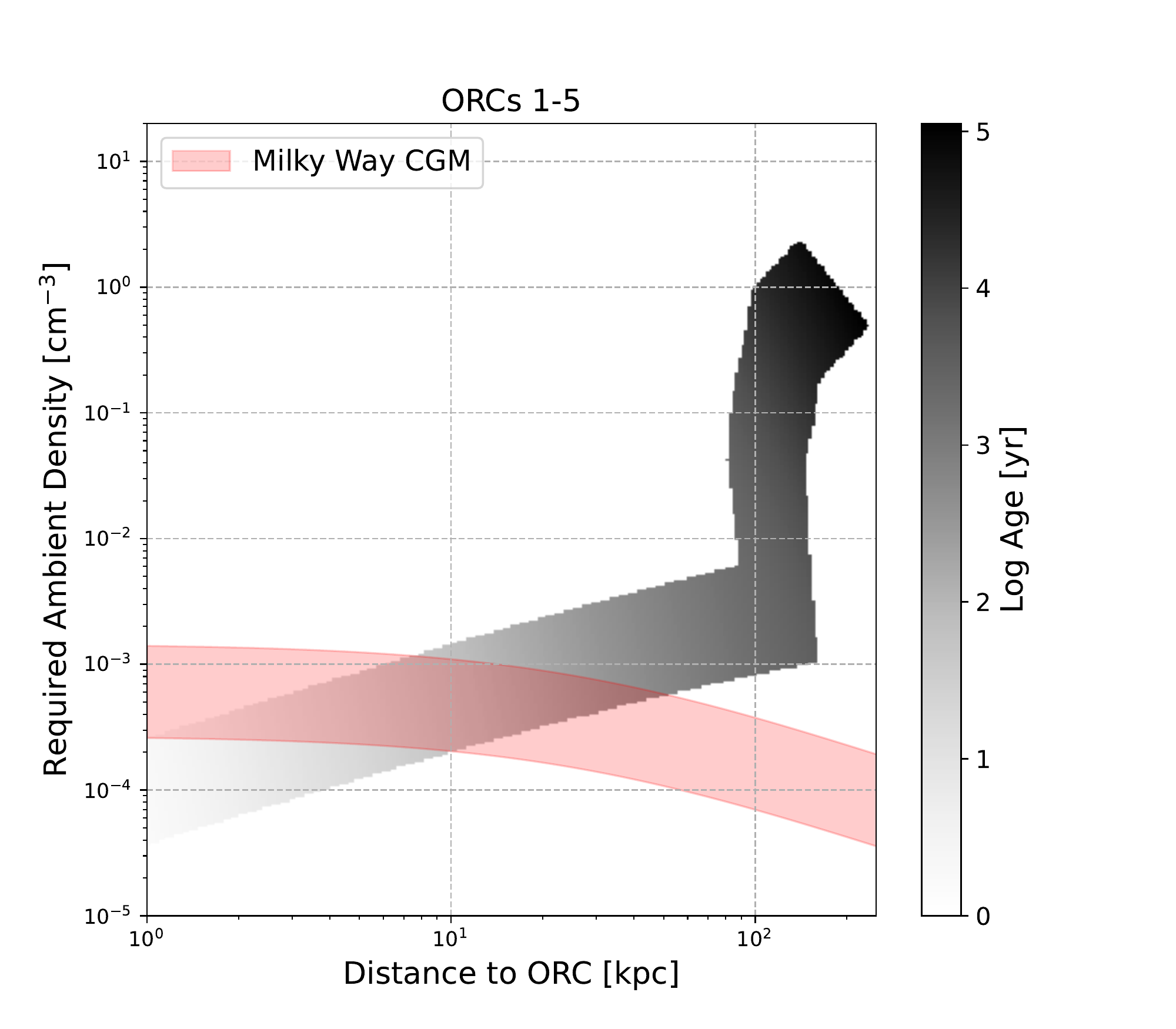}
        \caption{Ambient density needed to reproduce the 1GHz luminosities and diameters of ORCs 1-5 at a given distance from the Sun, assuming $10^{-3} \leq \epsilon_e \leq 10^{-2}$. From 0-100 kpc, the required density increases with distance as the ORC is in ejecta-dominated phase, while at 100-200 kpc, it is in the Sedov-Taylor phase, as explained in Section \ref{sec:requireddensitydistance}. Grey colorscale represents the logarithmic age of the SNR. The red shaded region represents the density profile of the Milky Way CGM used by Fang et al (2013) based on the models of Maller \& Bullock (2004), explained further in Section \ref{sec:cgm}.}
    \label{fig:dendistmw}
\end{figure}
We discuss here whether the ORCs 1-5 can be reproduced by our radio light curve models of SNRs at low densities. Since their distances are unconstrained, we calculate their luminosities and diameters at increasing distances from the Sun and check if the values are reproducible by our light curve models. Since ORCs 1-5 are similar in size and flux density, we use a median projected size of 80\asec and median flux density of 4 mJy for comparison with models. We then estimate the range of ambient densities versus distance within which the observed radio luminosities and diameters coincide with the model light curves for values of $\epsilon_e = 10^{-3} - 10^{-2}$. 

\subsubsection{Required ambient density and age range} \label{sec:requireddensitydistance}
Our results for plausible densities at various distances, are summarized in Figure \ref{fig:dendistmw}, and we refer the reader to the light curves in Figure \ref{fig:lum3panel} for reference. We also plot the age of the SNR, since older and larger SNRs are more frequent in a SNR population than younger and smaller ones.

Figure \ref{fig:dendistmw} suggests an upper limit of $\sim $200 kpc, which coincidentally is similar to the virial radius of the Milky Way \citep{Dehnen2006}. This can be understood from Figure \ref{fig:lum3panel}. As the distance increases, the ORCs are more luminous and have larger diameters for a given flux density and projected size. In our models, more luminous SNRs can be produced in a denser ambient medium, but this also results in smaller SNRs with shorter lifetimes as they decelerate and eventually merge more quickly with the ambient medium. In other words, more distant SNRs move towards the top right of Figure \ref{fig:lum3panel}, while model light curves in denser ambient media move towards the top left. Therefore, beyond a certain distance, the light curves and the observations cannot coincide, yielding an upper limit to their distance. We note that \cite{Omar2022} had similarly estimated that ORCs can be interpreted as intra-group medium SNRs going off at $0.1-3$ Mpc distances assuming a 500\,pc upper limit to the SNR diameter, though we find that these are unlikely to be farther than the virial radius of the Milky Way. 

Within 80-200\,kpc, the ORCs are consistent with a wide range of ambient densities spanning $\sim$$10^{-3}-2$ \cm, resulting in the parameter space extending vertically in Figure \ref{fig:dendistmw}. This is because at around 100 kpc, the ORCs coincide with the Sedov-Taylor part of the radio light curves as seen in Figure \ref{fig:lum3panel}, where luminosity decreases with diameter. With increasing density, the 100\,kpc luminosity and diameter is still encompassed by the Sedov-Taylor part of the light curve, but at progressively older ages. This trend continues up to densities of $\sim 2$ \cm, above which the predicted light curves are over-luminous and larger compared to the ORCs. 

At distances less than 80 kpc, the required ambient density range is less than $6 \times 10^{-3}$ \cm, and this range decreases with decreasing distance as seen in Figure \ref{fig:dendistmw}. This can again be understood from Figure \ref{fig:lum3panel} where the SNR luminosities and diameters for distances less than 80 kpc can only be encompassed by the ejecta-dominated part of the light curves for a smaller range of ambient densities than the Sedov-Taylor stage. The ages of the ORCs also decrease with decreasing distance, as shown by the progression of the gray colorscale with distance in Figure \ref{fig:dendistmw}. For example, ORCs at 4-10 kpc can only be reproduced by ambient densities $\sim(0.3-1) \times 10^{-3}$ cm$^{-3}$ and SNR ages of a few decades, and at 40 kpc, by an ambient density of $\sim 5 \times 10^{-4}$ cm$^{-3}$ and ages of a few hundred yrs. We note that the youngest SNRs known in the Local Group are SN1885A and G1.9+0.3, both Type Ia SNRs with ages of about 137 yrs and 100-140 yrs, respectively \citep{Fesen2007, Reynolds2008}.
\subsubsection{Comparison with Milky Way Circumgalactic Medium} \label{sec:cgm}
Is the required density parameter space consistent with densities expected in the Milky Way CGM? The CGM primarily consists of hot, low-density, ionized gas \citep{Spitzer1956, Tumlinson2017, Das2021}, and a variety of observations along multiple sightlines through the CGM indicate decreasing densities with distance, with typical values of 10$^{-3}$ \cm\ at few kpc from the Sun, and 10$^{-5}$ \cm\ out to the virial radius of the Milky Way \citep[see discussions in ][]{Das2019, Bhattacharyya2022}. For this paper, we compare our ORC density parameter space with the widely used CGM model of \cite{Maller2004}, which assumes the CGM as a hot, multi-phase adiabatic gas in hydrostatic equilibrium with the dark matter halo. We use the form of the model used by \cite{Fang2013}, who found consistency between the model and a variety of observations such as X-ray emission measures \citep[e.g.]{Henley2010, Gupta2012, Gupta2017}, pulsar dispersion measures along the LMC/SMC \citep{Crawford2001, Manchester2006}, and non-detection of neutral hydrogen in dwarf galaxies within 300 kpc of the Milky Way \citep{GP09}, while also being able to account for the missing baryonic mass of the Milky Way. We use the form of the density profile defined in Eq (1) of \cite{Fang2013}. The density range at a given distance is based on the assumption in \cite{Fang2013} that the integrated hot gas mass of the Milky Way CGM within the virial radius is in the plausible range of $0.5-2.7 \times 10^{11}$ \Msun. 

We show our results in Figure \ref{fig:dendistmw}. The density profile is somewhat anti-correlated with the required density versus distance constrained for the ORCs. Considering that most SNRs are found in their Sedov Taylor phase at ages of a few $\times 10^4$ yr, ORCs 1-5 as SNRs would most likely be evolving in ambient densities greater than 10$^{-2}$ \cm\ at a distance of $\sim 100-200$ kpc, where the parameter space in Figure \ref{fig:dendistmw} gets darker. However, these densities are several orders of magnitude higher than the CGM density profile in Figure \ref{fig:dendistmw} at those distances. While turbulence and inhomogeneities can produce overdense regions in the bulk CGM \citep{Oppen2016, Bhattacharyya2022}, these are not expected to be high enough to explain the required densities for ORCs.

One can also consider ORCs 1-5 to be SNRs at distances less than 50 kpc since their required density-distance parameter space overlaps with CGM densities in Figure \ref{fig:dendistmw}, but  such a scenario conflicts with the low SN rate within this volume and with the required ages of $\lesssim 640$ yr, as shown by the greyscale in Figure \ref{fig:dendistmw}. These ages are much smaller than their total lifetimes of nearly $10^5$ yr (Figure \ref{fig:lum3panel}), making it highly improbable that they will be observed at such an age. Given the total stellar halo mass of $1.4 \times 10^9$ M$_{\odot}$ \citep{Deason2019} with an average age of 10 Gyrs, the currently measured form of the SN Ia delay-time distribution of $(4 \times 10^{-13}\mathrm{yr^{-1} M_{\odot}^{-1}}) (t/\mathrm{Gyr})^{-1}$ \citep{Maoz2012} implies a SN Ia rate of $5.6 \times 10^{-5}$ yr$^{-1}$. Assuming these SNRs would be visible up to 10$^5$ yr based on Figure \ref{fig:lum3panel}, we expect only 6 SNRs in the halo. The stellar halo density profile is roughly $n_*(r) \sim r^{-3}$ \citep[e.g.][]{Slater2016, Iorio2018}, which means the number of stars at a given distance is $N_*(r) \sim \int_{r_0}^{r_1} n_*(r) d^3r \sim \mathrm{ln}(r_1/r_0)$. Thus, assuming a uniform distance distribution of SNe between $r_0=1$kpc and $r_1=260$kpc (roughly the virial radius), there is only a 19$\%$ probability of stars exploding within 50 kpc, meaning only $1-2$ SNRs are likely to exist within this distance today, and as mentioned earlier, these SNRs would most likely be older and larger SNRs. Indeed, some of the recently discovered and confirmed radio-faint halo SNRs by eROSITA are a few degrees across \citep{Churasov2021, Churasov2022, Becker2021}, much larger than the arcmin-sized ORCs.   

\subsubsection{Effects of model assumptions}
We checked if the results are robust to deviations from the fiducial values of model parameters in Section \ref{sec:modelassumptions}. The biggest effect comes from the assumption about $\epsilon_e$. A greater range of $\epsilon_e$ enlarges the allowed density-distance parameter space in Figure \ref{fig:dendistmw} because a larger range of luminosities are now allowed by the models in Figure \ref{fig:lum3panel}. For example, changing the $\epsilon_e=0.1$ leads to an upper limit of 350 kpc for ORCs 1-5, although such efficient electron acceleration efficiencies may not be physical in SNRs. The stopping condition, $M_s=1$ does not have an effect on the distance upper limit, but affects the maximum density allowed by the models as it leads to smaller final sizes for SNRs. E.g. if we use a stopping condition of $M_s = 5$, the maximum density in Figure \ref{fig:dendistmw} reduces from 1 \cm to 0.1 \cm. Changing $B_{low}$ within a factor of 2 of its fiducial value, i.e. $0.5-4$ $\mu$G does not change the maximum allowed distance of 200 kpc, and produces a minor shift, i.e. by few kpc, in the gray colorscale parameter space in Figure \ref{fig:dendistmw} to larger distances. This is because a larger $B_{low}$ increases the luminosity of the plateau portion of the light curve, producing a more luminous SNR at a given diameter. 

\subsubsection{Summary}
In summary, our S17 model implies that if ORCs 1-5 are SNRs, they must be going off within 200 kpc, and if they are interacting with the bulk CGM, they are most likely going off within 50 kpc in densities of $(0.2-1.2) \times 10^{-3}$ \cm. However, statistically their interpretation as SNRs is unlikely given their very young required ages and the low SN rate at the above distance range.
\begin{figure}
    \includegraphics[width=1.13\columnwidth]{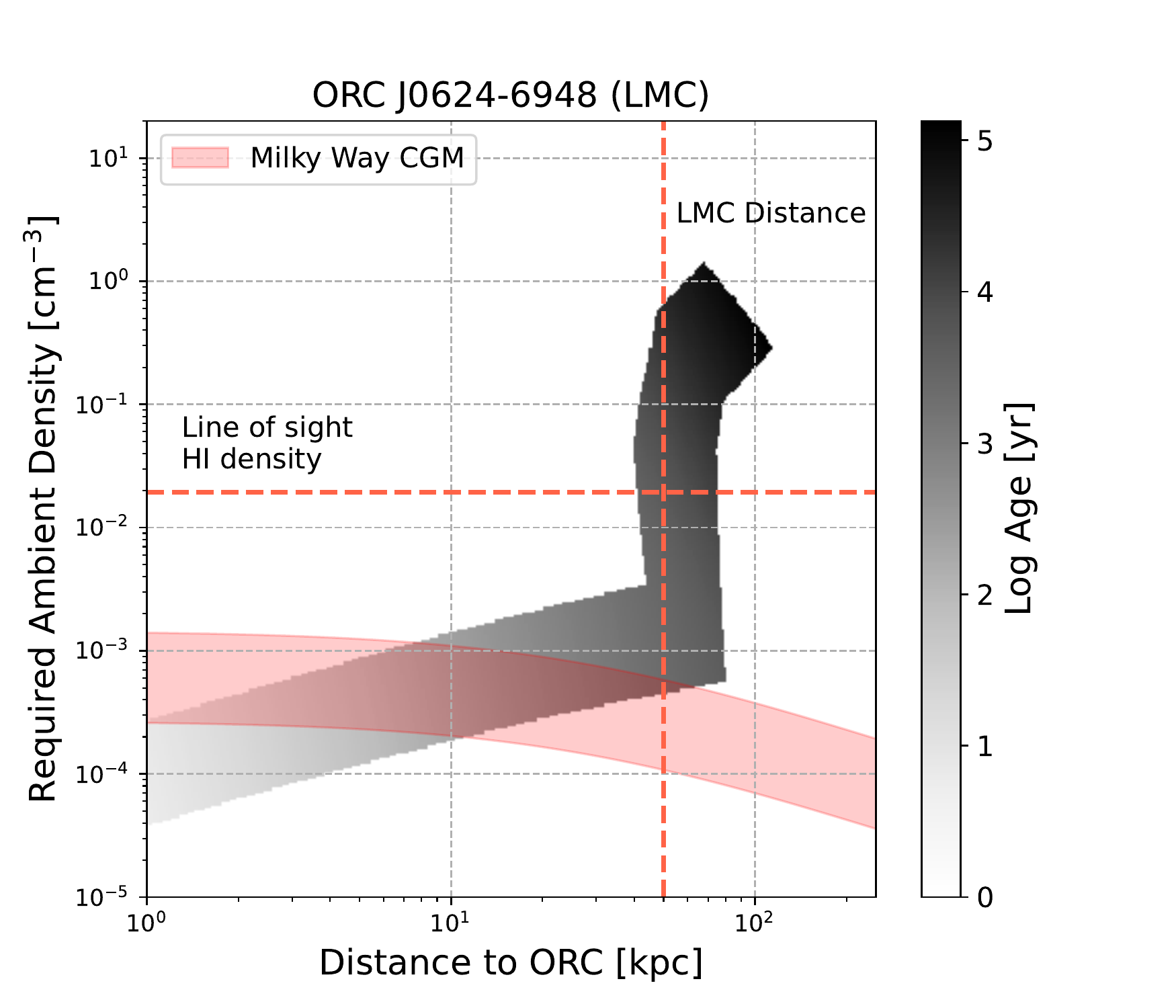}
    \caption{Same as Figure \ref{fig:dendistmw}, but for ORC \lmcorc. Vertical red line denotes the distance to LMC of 50 kpc, while horizontal red line denotes the line of sight HI density of 0.017 \cm. See Section \ref{sec:lmcsnr} for details.}
    \label{fig:dendistlmc}
\end{figure}

\subsection{Can ORC \lmcorc be an SNR?} \label{sec:lmcsnr} 
We apply the above analysis to \lmcorc as well, and show the results in Figure \ref{fig:dendistlmc}. Similar to ORCs 1-5, the distance to \lmcorc is unconstrained, but its close proximity to LMC also raises the possibility that it is located at the distance of LMC \citep[50 kpc,][]{LMCdistance}. We assume \lmcorc has a flux density of 11.16 mJy \citep[based on scaling the ASKAP 888 MHz flux density of 11.7 mJy to 1 GHz assuming a spectral index of $-0.4$ as constrained by][]{LMCORC} and use the measured angular size of 196$^{\prime\prime}$. We find a distance upper limit of about 100 kpc in Figure \ref{fig:dendistlmc}, about a factor of 2 less than ORCs 1-5. The density-distance parameter space behaves similar to the case of ORCs 1-5. The vertical region between $\sim 40-80$ kpc is where the observed luminosity and diameter at a given distance is reproduced by the Sedov Taylor part of the SNR light curves bounded by the values for $\epsilon_e = [10^{-3}, 10^{-2}$], while for distances less than 40 kpc, the observations are reproduced by the ejecta-dominated portion of the light curve (Figure \ref{fig:lum3panel}). 

We first consider the scenario where \lmcorc is evolving in the Milky Way CGM and happens to be projected along the LMC. Indeed there is some overlap between the CGM density profile and the ambient densities required by our model within a distance of 100 kpc, as shown in Figure \ref{fig:dendistlmc}. 
The maximum age that an ORC can have in this overlap region is about 3000 yr (based on the gray colorscale in Figure \ref{fig:dendistlmc}). 
Compared to ORCs 1-5 in Section \ref{sec:requireddensitydistance}, the allowed age range for \lmcorc in the CGM scenario is 3 times larger, but still much smaller than their lifetimes of $\sim 10^5$ yr. The CGM scenario is therefore more feasible for \lmcorc compared to ORCs 1-5, but still statistically unlikely due to the required young age. 

On the other hand, the interpretation of ORC \lmcorc as a Sedov-Taylor SNR in the LMC outskirts appears to be far more feasible based on Figure \ref{fig:dendistlmc}. At 50 kpc, the observed properties of \lmcorc can be reproduced by a wide range of densities ($6 \times 10^{-4} - 0.5$ \cm) and ages $\sim (0.2-2.6) \times 10^4$ yr. As discussed in \cite{LMCORC}, the best available constraint in the literature on the local environment comes from the HI4PI survey \citep{HI4PI}, which showed the presence of a diffuse HI cloud, about a degree across, along the line of sight of \lmcorc. The cloud is detected in emission integrated within the velocity range of 190-320 km/s, similar to the velocity component of the LMC shown in their Figure 9. With a column density of $6 \times 10^{19}$ cm$^{-2}$ after opacity correction, and taking the lateral extent of the cloud of 0.9 kpc as its line of sight depth, \cite{LMCORC} got an average density of 0.017 \cm, which falls well within the vertical density-distance region in our Figure \ref{fig:dendistlmc}. If the ORC is indeed evolving inside this cloud at 50 kpc, the ORC would correspond to an age of about 7000 years.

To summarize, the interpretation of \lmcorc as an SNR is more feasible compared to ORCs 1-5. Its interpretation as an ejecta-dominated SNR evolving in the CGM is more likely than ORCs 1-5 (though its required age is still much less than its expected lifetime at CGM densities) and its interpretation as a Sedov-Taylor SNR evolving in the diffuse HI environment in the LMC outskirts is consistent with the plausible model parameter space in Figure \ref{fig:dendistlmc}. The total stellar mass of the LMC is about an order an magnitude smaller than the Milky Way \citep{van2002, HZ09, Licquia2015}, so we expect the stellar halo mass to be similarly small. Thus, extending the SN rate calculation in Section \ref{sec:requireddensitydistance} to the LMC stellar halo implies $\lesssim$ 1 SNR, which is consistent with \lmcorc being the only ORC-like SNR discovered near the LMC in the EMU survey \citep{LMCORC}. It is possible therefore that some fraction of ORCs are SNRs like \lmcorc evolving in the low-density halo outside the star-forming disks of galaxies. Deep multi-wavelength follow-ups of existing ORCs, and continued deep radio-continuum surveys at high Galactic latitudes as well as the halo of nearby galaxies is warranted to further characterizing the nature of ORCs.

\section*{Acknowledgements}
SKS acknowledges support from the CCAPP Postdoctoral Fellowship, and useful discussions with Joy Bhattacharya, David Weinberg, and Sanskriti Das at the weekly OSU Galaxy/ISM meetings. LAL acknowledges support from the Heising-Simons Foundation. This research made use of \texttt{Astropy}, a community-developed core Python package for Astronomy \citep{2018AJ....156..123A, 2013A&A...558A..33A}, \texttt{NumPy} \citep{harris2020array}, \texttt{matplotlib}, a Python library for publication quality graphics \citep{Hunter:2007}, and NASA's Astrophysics Data System. The acknowledgements were compiled using the Astronomy Acknowledgement Generator.


\section*{Data Availability}
No new data were generated in support of this research. All analyzed data are publicly available in the citations mentioned in the text. The light curve model is hosted on \href{https://github.com/sks67/s17lc}{GitHub}.
 



\bibliographystyle{mnras}
\bibliography{new_main} 




\appendix
\section{Modified light curves in hot ISM} \label{sec:hotISM}
\begin{figure}
    \centering
    \includegraphics[width=\columnwidth]{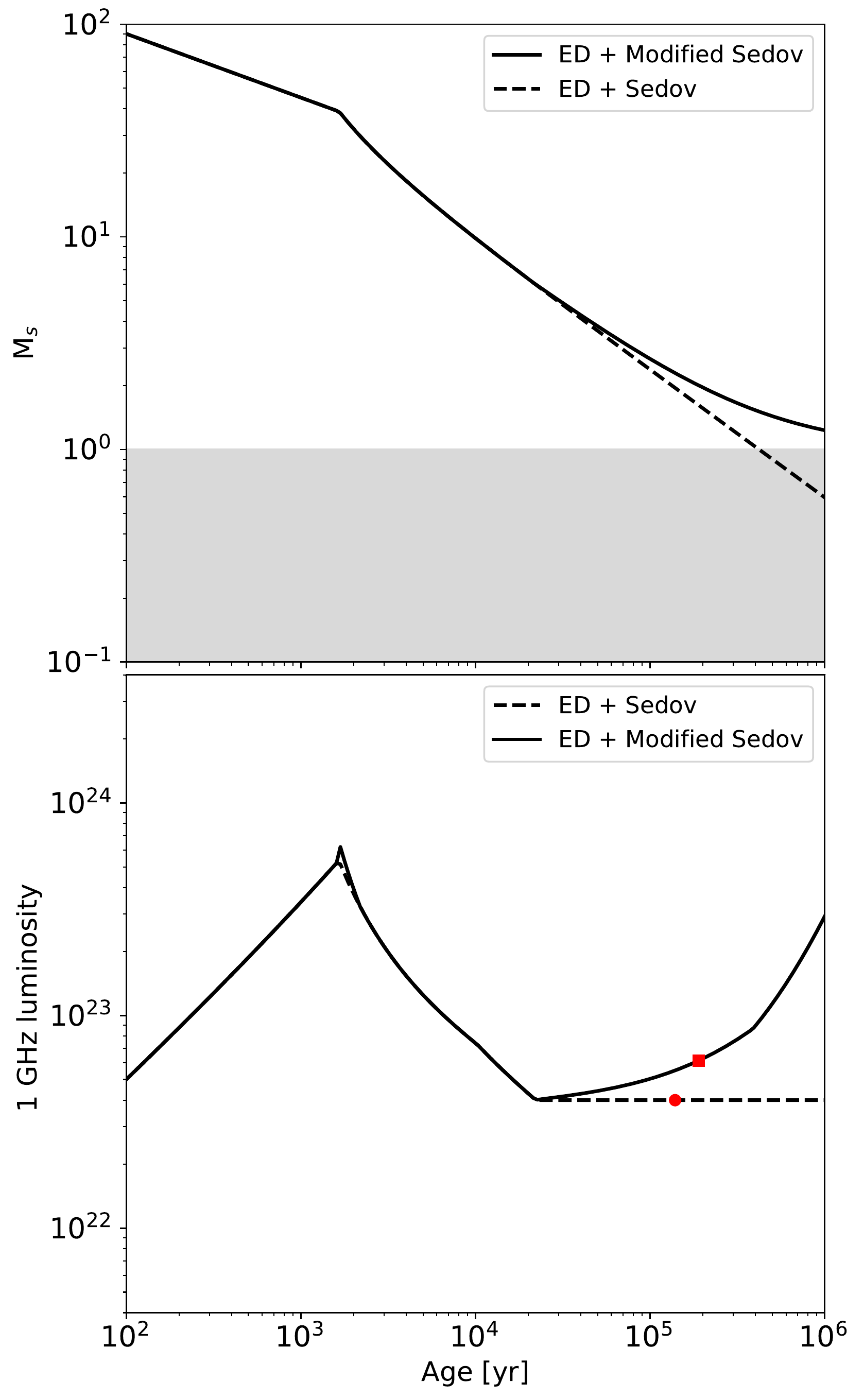}
    \caption{Sonic Mach number ($v_s/c_s$) and radio luminosity as a function of age for SNR for our default S17 model (dashed) and the S17 model including the deviation expected in the Sedov Taylor stage at high ambient temperatures (solid), based on Tang \& Wang (2005). Red circle and square denotes the age when $M_s = 2$ for the S17 light curve and modified light curve for hot ISM respectively. Case here is for $n_0 = 10^{-2}$ cm$^{-3}$, $T=10^6$ K.}
     \label{fig:hotISM}
\end{figure}
S17 model uses equations based on SNR evolution in the cold ISM ($T\lesssim 10^4$ K), where the SNR develops a cooling shell as it transitions from the Sedov to radiative phase. However, in the hot ($10^6$ K) ISM, the SNR never reaches the radiative phase as cooling becomes inefficient. \cite{Tang05} verified this with 1D simulations and found that at late times the shock radius and velocity evolution deviates from the Sedov-Taylor solution, and the shock velocity asymptotically reaches the sound speed of the ISM. The shock velocity in such a scenario was parameterized by \cite{Tang05} as
\begin{equation}
v_s(t) = c_s \left(\frac{t_c}{t} + 1\right)^{3/5}
\end{equation}
which gives the shock radius as $R_s(t) = \int_0^t v_s(t^{\prime}) dt^{\prime}$. Here $t_c$ is a characteristic time defined in \cite{Tang05} and corresponds to when the thermal energy added to the SNR by the swept-up ISM becomes twice the explosion energy. The expression goes as 
\begin{equation}
t_c = \left[ \left(\frac{2\xi}{3}\right)^5 \frac{E_{SN}}{\rho_0 c_s^5}\right]^{1/3}
\end{equation} 
where $\xi = 1.14$, $E_{SN}$ is the SN explosion energy, $\rho_0 = \mu n_0 m_p$ is the particle density in g cm$^{-3}$, and $c_s$ is the sound velocity of the plasma. 

We check the difference introduced by these results in the radio light curves in Figure \ref{fig:hotISM}. As shown in \cite{Tang05}, the solution for the shock velocity deviates from the Sedov solution and asymptotically reaches the sound speed for the corresponding ISM temperature at $M_s = 1$. The radio luminosity evolution closely follows the Sedov-Taylor solution we had in our S17 model, but starts to deviate around the same time the velocity solution deviates. Ultimately, the light curve appears to increase rapidly, as the shock velocity changes very slowly, while the shock emitting volume increases rapidly, similar to the cause for increasing luminosity in the ejecta-dominated phase. Both the visibility time and luminosity in our modified Sedov solution at $M_s = 2$ is within a factor of 2 of the corresponding values for our S17 model, which is a relatively small effect compared to the systematic uncertainties introduced by other parameters in our model (e.g. $\epsilon_e$) that affect the luminosity by order of magnitude. 

While we could adopt this solution into our model for the present study, we refrain from doing that until these have been tested properly with observed SNRs, particularly since it predicts a rise in luminosity at late times. Additionally, these results are borne of 1D simulations, which do not necessarily capture physics at higher dimensions that can affect the shock evolution, such as fluid instabilities that enhance mixing of the hot and cold phases \citep{Gentry2019}.

\bsp	
\label{lastpage}
\end{document}